\documentclass[a4paper,11pt]{article}
\usepackage{pos}

\renewcommand{\logo}{\relax} 

\input epsf

\title{Holographic Quantum Foam: Theoretical Underpinnings and Observational Evidence}
 \ShortTitle{Holographic Quantum Foam}

\author*[a]{Eric Steinbring}

\affiliation[a]{Herzberg Astronomy and Astrophysics, National Research Council Canada,\\
  5071 West Saanich Road, Victoria, British Columbia, Canada}

\emailAdd{Eric.Steinbring@nrc-cnrc.gc.ca}

\author*[b]{Y. Jack Ng}

\affiliation[b]{Department of Physics and Astronomy, University of North Carolina,\\
  120 E. Cameron Avenue, Chapel Hill, North Carolina 27599, USA}
  
\emailAdd{yjng@physics.unc.edu}  

\abstract{Spacetime is foamy due to quantum fluctuations. Various gedanken experiments show that distances fluctuate by amounts consistent with the holographic principle, hence the name "holographic quantum foam" (HQF). One important prediction of HQF is that necessarily there exists a dark sector in the universe.  The resulting cosmology is found (at least qualitatively) to be consistent with observations.  Interestingly the quanta of the dark sector are found not to obey the familiar (fermionic or bosonic) statistics, but the exotic statistics known as infinite statistics (or quantum Boltzmann statistics). The most important challenge now is to check if HQF is consistent with experiments/observations. One way is to look for observational evidence of blurred distant point-sources due to physics at the Planck scale. For over two decades it has been debated whether those tiny inherent uncertainties in time and path-length can accumulate in transiting electromagnetic wavefronts from quasars and Gamma-Ray Bursts (GRBs). But a recent event is special: GRB221009A was extremely bright and energetic. That allowed follow-up across the whole spectrum from the optical/near-infrared through to X-rays, and including the highest-ever-recorded energy gamma-rays; all consistent with blurring by HQF. Those data, and a calculation of the HQF-widened point-spread function (PSF) for real telescopes viewing a GRB are presented.} 

\FullConference{Proceedings of the Corfu Summer Institute 2025 "School and Workshops on Elementary Particle Physics and Gravity" (CORFU2025)\\
27 April - 28 September, 2025\\
Corfu, Greece\\}

\begin{document}
\maketitle

\section{Introduction}

The idea of spacetime foam is believed to be due to John Wheeler \citep [1957; see][for a recent review]{Carlip2023}.  Probed at small scales, spacetime appears to be very complicated -- something akin in complexity to a turbulent froth which he dubbed spacetime foam. Many physicists believe the foaminess is due to quantum fluctuations of spacetime, hence spacetime foam is a.k.a. quantum foam. But how large are the quantum fluctuations? How foamy is spacetime?  More specifically, let us start with the question: What is the uncertainty $\delta l$ in the measurement of distance $l$? We will see that $\delta l \gtrsim l^{1-\alpha} l_P^{\alpha}$ on the average, with the Planck length $l_P \equiv \sqrt{\frac{\hbar G}{c^3}} \sim 10^{-33}$ cm providing the intrinsic length scale in quantum gravity, and the parameter $\alpha \sim 1$ characterizing different spacetime foam models. 

This talk consists of two parts: the first part (sections 2 and 3) has to do with the theoretical underpinnings of HQF. In section 2, we use four very different arguments to show that $\alpha = 2/3$, which is shown to be consistent with the holographic principle \citep{Susskind1995, 't Hooft1993}. In section 3, we employ a simple argument based on statistical mechanics to show that the universe cannot contain ordinary matter only. The theory of HQF leads to a predictive framework that ties together the nature of dark energy and quantum foam. We further show that the quanta of dark energy necessarily obey not the familiar statistics, but the so-called infinite statistics (sometimes known as the quantum Boltzmann statistics). Then we discuss briefly the holographic foam cosmology incorporating also dark matter, the quanta of which likewise obey infinite statistics. Finally, we use the deep similarities between the spacetime foam phase of strong gravity and turbulence to argue that HQF is consistent with a cosmic inflation in the early universe.  

The second part of this talk (sections 4 and 5) deals with observational evidence of HQF.  
Since the time of Wheeler's proposal of a foamy spacetime, observational tests of metric graininess considered how light propagation from sources a distance $L$ away might be measurably affected via transit through such a foam. And whether photons are randomly ``jiggled'' in some way throughout their long paths to an observer here on Earth has been searched for intensely over the last two decades or so. Lorentz invariance could be violated, via photons given a slightly broadened energy dispersion $\delta E$ scaling as $L/c$ \citep[as reviewed in][]{Amelino-Camelia2013}. Another independent experimental avenue would be to look for images of those sources being instead visibly blurred by accumulating microscopic distance fluctuations $\pm \delta l$ proportional to Planck length $l_{\rm P} \sim {10}^{-35}$ m (or equivalently, timescale $t_{\rm P}\sim 10^{-44}~{\rm s}$) in their electromagnetic wavefronts. 

Evidence of either effect would not only demonstrate that spacetime is foamy, but could help differentiate between predictions of various quantum-gravity (QG) theories. For example, if the blurring effect occurs, wavefront phase degradation at observed wavelength $\lambda$ should be the sum of each small phase perturbation $\Delta \phi = 2\pi \delta l/\lambda$, accumulated along photon path of length $L$ as $$\Delta\phi_0=2\pi a_0 {l_{\rm P}^{\alpha}\over{\lambda}}L^{1 - \alpha}\eqno(1)$$for $a_0\sim 1$ and $\alpha$ specifying the QG model \citep{Ng2003}: $1/2$ is for a simple random walk, which would be a very strong effect; $2/3$ results in a much weaker blur, but as will be discussed in sections 2 and 3, it is consistent with the holographic principle and expectations of information theory related to the surface of a black hole, and is of particular interest; no phase degradation at all would occur for $\alpha=1$, and so leave perfectly pristine images of any distant pointsource at all wavelengths. 

\begin{figure}
\includegraphics[scale=0.67]{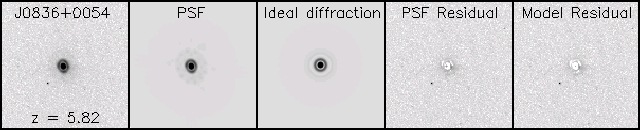}
\caption{Starting from left; reproduced from \cite{Steinbring2007}: quasar J0836+0054 image with HST; the intrinsic PSF, which includes known instrumental aberrations; and the ideal telescope diffraction pattern (middle panel). This showed a faint PSF residual, with some evidence of Planck-scale induced blurring via a drop in encircled photon flux within the central diffraction spike, when modeled for expected size including foam (right panel). As high-$z$ quasars can be partially resolved, this cannot be called a detection of foam, but did set limits on it.}
\label{figure_quasar_images}
\end{figure}

Unfortunately, this latter observational approach requires distant, bright pointlike objects to test, as the Planck-scale-induced phase error is compared directly with the intrinsic point-spread function (PSF) of the telescope: how sharp the image should be. And in optical light that effect is only just-comparable to the diffraction-limit of the {\it Hubble Space Telescope} (HST). This still allowed limits to be put on it, proving it to be weaker than $\alpha\leq 0.65$ via images of distant galaxies \citep{Lieu2003, Ragazzoni2003} and active-galactic nucleii \citep[AGNs;][]{Steinbring2007, Christiansen2011, Perlman2011, Tamburini2011}. Doubly unfortunate though, quasars and other AGNs are known to be up to kiloparsecs in diameter, and so where observed at high redshift can have angular sizes on the sky also comparable to equation 1, confusing the issue for the smallest-discernible effect, and leading to an inconclusive result: see Figure~\ref{figure_quasar_images}. 

Although far from diffraction limited, more recent observations of gamma-ray bursts (GRBs) with {\it Fermi} since its launch in 2008 give more stringent constraints; blurring at gamma-ray energies could scatter some photons across the whole sky! And, thankfully, GRBs are much more compact than quasars, with emission regions less than a parsec across; so a detection can be conclusive. It is still consistent with equation 1 and HQF ($\alpha=0.667$) at high energy after accounting for observational effects with X-ray and gamma-ray telescopes. And intriguingly, blurred GRBs observed at the highest energies reach something like a universal average PSF for any telescope, regardless of its native resolution, which is akin to how atmospheric scattering affects optical images of pointsources from the ground. Most importantly, there was a recent one that was particularly energetic, and in fact the brightest ever seen: GRB221009A. It was observed in the optical through to the highest-observed energies for any GRB, and those results are summarized here.

\section{Various Ways to Calculate $\delta l$}

In the following four subsections we will consider four very different ways to calculate $\delta l$, the accuracy with which distance $l$ can be measured.  More detailed arguments on some technical parts can be found in \cite{Ng2025} and references contained therein. For convenience, very often we will use units such that  $c = 1, \hbar = 1$, Boltzmann constant $k_B = 1$, and we will feel free to drop multiplicative factors of order 1.

\subsection{Method 1: A Simple Gedanken Experiment}

To measure the distance $l$ between two points, let us put a clock (of mass $m$) at one end and a mirror at the other end. By sending a light signal from the clock to the mirror in a timing experiment we can determine $l$ [\citealp{Ng1994}, \citealp{Karolyhazy1966}]. 
Let us concentrate on the uncertainties in the positions of the clock when the light signal leaves (at time $t = 0$) and then returns to the clock (at time $t = 2l/c$).  Using Heisenberg's uncertainty principle we find $\delta l \left(\frac{2l}{c}\right) = \delta l + \frac{2l}{c} \frac{1}{m} \frac{\hbar}{2 \delta l}$ yielding $\delta l^2 \gtrsim \frac{\hbar l}{mc}$. But $\delta l$ cannot be made arbitrarily small by using a very massive clock as that would disturb the curvature according to general relativity.  The error $\delta l$ caused by the curvature can be estimated from the Schwarzschild solution giving $ \delta l \gtrsim \frac{Gm}{c^2} $.  It follows that the combined error/uncertainty in the length messurement is $$\delta l \gtrsim l^{1/3} l_P^{2/3}. \eqno(2)$$

\subsection {Method 2: Applying the Holographic Principle}

The holographic principle essentially stipulates that the Universe which we perceive to have 3 spatial dimensions can be encoded on a 2-dimensional surface, like a hologram \citep{'t Hooft1993, Susskind1995}. In other words, the maximum amount of information contained in a region is bounded by its surface area: i.e., the number of degrees of freedom is bounded by its surface area in units of $l_P^2$.

Let us consider a (big) cube with side $l$ and partition it into (small) cubes (with average size $= (\delta l)^3$) as small as physical laws allow; so intuitively one degree of freedom is associated with each small cube. It follows that the number of degrees of freedom inside $l^3$ is equal to the number of small cubes $\left(\frac{l}{\delta l}\right)^3$ which is, order of magnitudewise, bounded by $\frac{l^2}{l_P^2}$ by the holographic principle yielding $\delta l \gtrsim l^{1/3} l_P^{2/3}$, in agreement with equation 2.  And naturally the quantum foam model corresponding to $\delta l \gtrsim l^{1/3} l_P^{2/3}$ is now known as the holographic quantum foam (HQF) model .

\subsection {Method 3: Using the causal-set approach to quantum gravity}

In this subsection we will rederive the magnitudes of $\delta l$ by using causal-set theory. The causal-set theory \citep{Sorkin1997} stipulates that continuous geometries in classical gravity should be replaced by ``causal-sets", the discrete substratum of spacetime. In the framework of the causal-set theory, the fluctuation in the number of elements $N$ making up the set is of the Poisson type, i.e., $\delta N \sim \sqrt{N}$. For a causal set, the spacetime volume $V_{st}$ becomes $l_P^4 N$. It follows that $\delta V_{st} \sim G \sqrt{V_{st}}$. Let us consider a spherical volume of radius $l$ over the amount of time $T = 2l/c$ it takes light to cross the volume. We want to find the minimum of $\delta l$; so $\delta V_{st} \sim T (\delta l)^3 \sim l (\delta l)^3$. With the help of $\sqrt{V_{st}} \sim l^2$, we recover $\delta l \gtrsim (l l_P^2)^{1/3}$.


\subsection {Method 4: Mapping the Geometry of Spacetime}

Consider using a global positioning system by filling the space with a swarm of clocks, exchanging signals with the other clocks and measuring the signals' time of arrival.  The question we want to address is how accurately can these clocks (of total mass $M$) map out a volume of space-time with radius $l$ over time $l/c$ it takes light to cross the volume \citep{Lloyd2004}. Since the process of mapping the geometry of spacetime is a kind of computational operation, we can apply the Margolus-Levitin theorem \citep{Margolus1998} which stipulates that the rate of operations is bounded by the energy $E$ which is available to perform the computation.  Hence the number of operations is bounded by $(E/ \hbar) \times (l/c)= \frac{Mc^2}{\hbar} \frac{l}{c}$. Imposing the condition $M < \frac{lc^2}{G}$ for the space under consideration not to collapse into a black hole we get the number of operations or events (i.e., number of spacetime ``cells'') being bounded by $< l^2 \frac{c^3}{\hbar G} = {\frac {l^2}{\l_P^2}}$. For maximum spatial resolution, each clock ticks only once. Then it follows that each ``cell'' occupies a spatial volume of $\frac{l^3}{l^2/l_P^2} = l l_P^2$ so that the average spatial separation of ``cells'' is $l^{1/3} l_P^{2/3}$. Thus heuristically, we have again arrived at the result $\delta l \gtrsim l^{1/3} l_P^{2/3}$. 

It will prove to be useful at this point to note that maximum spatial resolution requires maximum energy density $\rho \sim \frac{3}{8 \pi} (l l_P)^{-2}$, yielding the total number of bits of information being of the order $ l^2/l_P^2$.

\section{Holographic Foam Cosmology}

In this section we will show that HQF gives rise to an exciting cosmology which automatically accomodates (actually requires) a dark (nonluminous) sector which is very different from the familiar particles. Then we proceed to briefly discuss some aspects of dark energy, dark matter, and cosmic inflation in the early universe.

\subsection{Prediction of a Dark Sector in the Universe}

Let us generalize the quantum foam discussion for a static spacetime region (with low spatial curvature) to the case of an expanding universe. We will start by assuming the Universe (of size $l = R_H$, the Hubble horizon) has only ordinary matter. According to the statistical mechanics for ordinary matter at temperature $T$, (and in volume $ l^3$) energy scales as $E \sim l^3 T^4$ and entropy goes as $S \sim l^3 T^3$.  Black hole physics can be invoked to require $E \lesssim \frac{l}{G} = \frac{l}{l_P^2}$. Then we expect that entropy $S$ (and hence also the number of bits or the number of degrees of freedom) is bounded by $\lesssim (l / l_P)^{3/2}$. 

We can repeat verbatim our earlier argument to conclude that, if only ordinary matter exists, $\delta l \gtrsim \left( \frac{l^3}{(l / l_P)^{3/2}} \right) ^{1/3} =  l^{1/2} l_P^{1/2}$ which is much greater than $l^{1/3} l_P^{2/3}$ for the case of the HQF model. In other words, ordinary matter contains only an amount of information dense enough to map out spacetime at a level with much coarser spatial resolution. Thus, there must be other kinds of matter/energy with which the Universe can map out its spacetime geometry to a finer spatial accuracy than is possible with the use of conventional ordinary matter. We conclude that there must be a dark sector in the Universe \citep{Arzano2007}! (In principle, one can draw this conclusion, independent of recent observations of dark energy and dark matter.)

In passing, we can mention that the quantum foam model corresponding to $\delta l \gtrsim  l^{1/2} l_P^{1/2}$ is appropriately called the random walk foam model.  This corrresponds to the case if only ordinary matter exists in the Universe. For the different spacetime foam models, we can write $\delta l \gtrsim l^{1 - \alpha} l_P^{\alpha}$, with the values of $\alpha = 1/2, 2/3, 1$ corresponding respectively to the random walk model, the holographic model, and the minute Planck length fluctuation model.

\subsection{Dark Energy}

As shown in subsection 2.4, for the current expanding universe, holographic foam cosmology is characterized by these two main features: (1) critical cosmic energy $\rho = \frac{3}{8 \pi} \left(\frac{H}{l_P}\right)^2 \sim (R_H l_P)^{-2}$; (2) the Universe contains $I \sim (R_H/l_P)^2$ bits of information, where $H, R_H$ stand for Hubble parameter, radius respectively. Thus, the average energy carried by each bit is $\rho R_H^3/I \sim R_H^{-1}$. In othere words, dark energy (DE) acts like a dynamical cosmological constant $\Lambda \sim 3 H^2$, i.e., DE is composed of very long wavelength ``particles''.  (Note the quotation marks around the word `particles' since they hardly qualify to be called particles.) Obviously we would like to know how different these ``particles'' are (from ordinary particles). 

Consider $N \sim (R_H/l_P)^2$ such ``particles'' obeying Boltzmann statistics in volume $V \sim R_H^3$ at temperature $T \sim R_H^{-1}$. The partition function $Z_N = (N!)^{-1} (V / \lambda^3)^N$ yields entropy of the system being $S = N [ln (V / N \lambda^3) + 5/2]$ with thermal wavelength $\lambda \sim T^{-1}$. But $V \sim \lambda^3$, so $S$ becomes negative unless $N \sim 1$ which is equally nonsensical. The only sensible solution is that the $N$ inside the log in $S$, i.e, the Gibbs factor $(N!)^{-1}$ in $Z_N$, must be absent, implying that the N ``particles'' are distinguishable!  Then $S = N[ln (V/ \lambda^3) + 3/2]$ is positive, as required \citep{Ng2007}.

Now the only known consistent statistics in greater than 2 space dimensions without the Gibbs factor is the quantum Boltzmann statistics, a.k.a. the infinite statistics \citep{Greenberg1990}. So the only logical proposal is that the ``particles'' constituting dark energy obey infinite statistics, rather than the familiar Fermi or Bose statistics.  This is the overriding difference between DE and conventional matter.

For the sake of completeness, we conclude this subsection by mentioning a few characteristic properties of infinite statistics. This ``exotic'' statistics is given by the Heisenberg algebra $a_k a^{\dagger}_l = \delta_{kl}$. Any two states obtained by acting on the vacuum $|0>$ with creation operators in different orders are orthogonal to each other, implying that particles obeying infinite statatistics are virtually distinguishable. The partition function is given by $Z = \Sigma e^{- \beta H}$, with NO Gibbs factor. In infinite statistics, all representations of the particle permutation group can occur. Furthermore, theories of particles obeying infinite statistics are non-local; for example, the number operator $n_i = a_i^{\dagger} a_i + \sum_k a_k^{\dagger} a_i^{\dagger} a_i a_k + \sum_l \sum_k a_l^{\dagger} a_k^{\dagger} a_i^{\dagger} a_i a_k a_l + ...$, and Hamiltonian are both nonlocal and nonpolynomial in the field operators. But the TCP theorem and cluster decomposition still hold; and quantum field theories with infinite statistics remain unitary.

\subsection {Dark Matter}

We start with the observation that Milgrom's Modified Newtonian Dynamics (MoND) \citep{Milgrom1983} can account for the flat galactic curves and the Tully-Fisher relation at the galactic scale whereas the cold dark matter (CDM) paradigm cannot easily do.  On the other hand, there are problems with MoND at the cluster and cosmological scales, where apparently CDM works much better.  A rather obvious question is: Can we reconcile the dark matter and MoND approaches by introducing a new concept of ``MoNDian dark matter (MDM)" in a scheme such that the MDM looks like CDM at cluster and cosmological scales, but it behaves like MoND at the galactic scale? Ho \citep{Ho2012} and his collaborators take that challenge by proposing the MDM approach in which Milgrom's scaling can be derived from holographic foam cosmology (HFC)! They show that MoND is simply a phenomenological consequence of HQF as the dynamical cosmological constant in HFC automatically gives rise to a critical acceleration parameter of the same magnitude as the one Milgrom had to put in by hand in MoND. Furthermore there is a surprising connection between the proposal by Ho et al. and an effective gravitational Born-Infeld description of the MOND-like phenomenology of the dark matter quanta. It is also shown that these unusual quanta of dark matter must obey the crucial property of infinite statistics. Thus, MDM has to be described as an essentially non-local theory for such infinite statistics quanta. Such a theory would be fundamentally quantum gravitational and thus distinguished from the usual phenomenological models of dark matter.

In the MDM scenario, at the galactic scale, a connection among dark matter ($M'$), ordinary matter ($M$) and dark energy (characterized by cosmological constant $\Lambda$) given by $M' \sim (\sqrt{\Lambda}/G)^{1/2}M^{1/2}r$ is found. Ho and collaborators successfully fit rotation curves for 30 local spiral galaxies; they also find that MDM fares well at cluster scales.

\subsection {Quantum Foam, Turbulence and Inflation in the Early Universe}

There are deep similarities between the problem of quantum gravity and turbulence \citep{Jejjala2008}. The connection between these seemingly disparate fields is provided by the role of diffeomorphism symmetry in classical gravity and the volume preserving diffeomorphisms of classical fluid dynamics. These similarities can be used to show that Kolmogorov's ``two-thirds law'' in turbulent fluid dynamics \citep{Kolmogorov1941} can be satisfied in the context of HQF. Thus Wheeler's picture of spacetime foam is correct: quantum foam (of the holographic type) IS indeed a turbulent froth!  

Holographic quantum foam can also help in shedding light on the cosmic inflation in the early universe. After all, one of the HFC's main features , that the cosmic energy is of critical density, is a hallmark of the inflationary universe paradigm.  But what led to the brief inflation? It has been proposed \citep{Ng2021} that it was due to the onset of cosmic turbulence in the early universe. Furthemore, HQF can provide a natural mechanism for the universe to transit ``gracefully`` from a turbulent phase to a laminar phase to end inflation in the process --- by virtue of the nonlocality property enjoyed by the quanta of spacetime foam.

\section{How Telescopes Can See Quantum Foam}

Having thoroughly mapped out the theoretical landscape of quantum foam, and enumerated its potentially observable properties, we return to one particularly promising avenue for its definitive detection. That is to sense blurring of distance pointsources. 

In early attempts to detect spacetime foam, quasars were used as pointlike sources to look for degradation of wavefront coherence of wavefronts in their long transit through spacetime. But GRBs are much better targets for probing this than quasars. Not only do GRBs provide higher-energy photons and more brightly (but temporarily) outshine their host galaxy, the central emission region of all those photons is ``tiny'': their initial flashes with which they are detected are only minutes to hours long, or perhaps in extreme cases days in duration; so all occurs within light-travel distances much less than a parsec. Scattering of these emitted high-energy photons by physics other than via quantum foam can still be looked for, i.e. from weak intragalactic or Galactic magnetic fields; but only detected photons contribute to a blurred image, so a spatially-uniform random process like absorption via extragalactic background light is not relevant, even at the highest possible energies. What matters is that the bulk of their flux across the whole electromagnetic spectrum started out as pointlike, as viewed from Earth. Quasars, by comparison, are huge: their emission comes instead from partially dust-enshrouded accretion disks and extended jets, embedded within large galaxies of comparable brightness, which can be $\sim 250~{\rm kpc}$ across. 

\subsection{Gamma-Ray Bursts Observed at All Energies}

The physical compactness of GRBs can overcome the instrumental PSF limitations of X-ray and gamma-ray telescopes, which are much larger than diffraction, that is $\lambda/D<<1$, where $D$ is the telescope diameter $\sim 1~{\rm m}$. But that is in tension with $\alpha=0.667$ and equation 1, as any GRB emitting photons of over 100 MeV - even one with a modest redshift of only $z=0.10$ - each wavefront phase dispersion $\Delta \phi_0$ would be over 1 radian. Interpretting this angle to be equivalent to a deflected ray, if all wavefronts are approximately that decohered, the source distribution of photons on the sky would be nearly uniform: no GRBs would be identifiable with a galaxy host!

Thousands of GRBs have been detected with the {\it Fermi} wide-angle Gamma-ray Burst Monitor (GBM); to which the spacecraft later re-pointed, shifting by what is called a ``roll angle'' of many degrees to nearly half-way across the sky (i.e. possibly a large number, but always less than the zenith angle to the celestial pole coordinates); then sensing most photons within some ``error radius'' degrees across; in order to view with the Large Area Telescope (LAT) and so produce an image, sufficient to localize at its native ``resolution'': a few arcminutes. Hundreds of these sources have been followed up on, with spectroscopic redshifts up to $z\approx4$ or so, and positions refined to less than an arscecond on the sky, via space and ground-based optical/near-infrared observations.

An instrumental PSF model accounting for these three reported measurements was developed by Steinbring \citep{Steinbring2015, Steinbring2016} and resolves the tension by recognizing that a random process can allow blurring to be either more or less than equation 1, so $\alpha$ need not be larger, as proposed \cite{Perlman2015, Ng2022}. This provides a mean angle-of-arrival of all detected photons which is the accumulated long-exposure image generated from those, and includes only those photons scattered less than the telescope field of view (FoV) in addition to telescope diffraction, and down through to some absolute minimum deflection which could be attributable to the Planck scale; so it remains applicable at all wavelengths.

\subsection{Derivation of the Quantum-Foam-Induced Point-Spread Function}

Working towards higher energies (from red to blue wavelengths), consider where photons can successively be scattered as much as $2\pi$, and so are no longer localizable by a telescope because those could have arrived from any angle on the sky. Notice the maximal blurring can be only $$\Delta\phi_{\rm max} = \Delta\phi_{\rm los} + \Delta \phi_z = 2\pi a_0 {l_{\rm P}^{\alpha}\over{\lambda}}\Big{\{}\int_0^z L^{1 - \alpha} {\rm d}z + {{(1-\alpha)c}\over{H_0 q_0}}\times \int_0^z (1+z) L^{-\alpha} \Big{[}1 - {{1 - q_0}\over{\sqrt{1 + 2 q_0 z}}}\Big{]} {\rm d}z \Big{\}}$$ $$ = (1+z)\Delta\phi_0, ~~~~~~~~~~~~~~~~~~~~~~~~~~~~~~~~~~~~~~~~~~~~~~~~~~~~~~~~~~~~~\eqno(3)$$where $L = ({c/{H_0 q_0^2}})[q_0 z - (1 - q_0)(\sqrt{1 + 2 q_0 z} - 1)]/(1+z)$ is the comoving distance and deceleration parameter $q_0={{\Omega_0}/{2}} - {{\Lambda c^2}/{3 H_0^2}}$ assumes a standard $\Lambda$CDM cosmology; values of $\Omega_\Lambda=0.7$, $\Omega_{\rm M}=0.3$ and $H_0=70~{\rm km}~{\rm s}^{-1}~{\rm Mpc}^{-1}$ will be used throughout, as in \cite{Steinbring2007}. This gets stronger with bluer light, of course; here, $\Delta\phi_{\rm los}$ includes waves propagating from any point along the line of sight, and $\Delta \phi_z$ are only those redshifted to the observer. The ratio between the greatest and the least-possible effect is always $\Delta \phi_{\rm max} / \Delta \phi_{\rm P}=(1 + z) a_0 (L/l_{\rm P})^{1 - \alpha}$, with no dependence on $\lambda$, and where $\Delta \phi_{\rm P} = 2\pi{{l_{\rm P}}\over{\lambda}}$ is the Planck-length minimum. Thus, at the $\lambda$ where these maximally-scattered photons cross the horizon, a long exposure produces an image averaging only the detectable phase dispersions up to that, and if those have a distribution with amplitude ${\Delta \phi}~\sigma (\Delta \phi) = 1-A \log({{\Delta \phi}/{\Delta \phi_{\rm P}}})$, $${1\over{A}} \int \Delta \phi ~\sigma (\Delta \phi) ~{\rm d}{\Delta \phi} = (1 + z) \Delta \phi_0, \eqno(4)$$for $A = 1/\log{[(1 + z) a_0 (L/l_{\rm P})^{1 - \alpha}]}$, which is constant for all $\lambda$. So, some photons might be dispersed by $\Delta\phi_{\rm max}$ but the average blurring at any wavelength is always less, and redward of this ``horizon-crossing'' limit it should maintain a decaying power-law wavelength dependence. And blueward of that, any point source viewed by a telescope allows sensitivity only to dispersions smaller than an opening angle $\theta$ which is less than its FoV (and the horizon); an angle ``Theta'' of $1^\circ$ is a useful benchmarch, for example. With $\theta \leq 2\pi$ and $A>0$, this implies a PSF mean width \citep{Steinbring2015} $$\Phi = \Phi_R + \Phi_\theta = R\Big{(}{{\lambda}\over{D}}\Big{)}^\rho + \int_0^\theta \Delta \phi ~\sigma (\Delta \phi) ~{\rm d}{\Delta \phi} ~~~~~~~~~~~~~~~~~~~~~~~~~~~~~~~~~$$ $$~~~~~~~~~~~~~~~~~~~~~~~~~~~~~~~~~~~~~~~~~ = A R \Big{(}{\lambda\over{D}}\Big{)}^\rho \Big{[} 1 + \log{\Big{(}{{2\pi l_{\rm P} D^\rho}\over{R \lambda^{\rho+1}}}\Big{)}}\Big{]} + \theta \Big{\{} 1 + A\Big{[} 1 + \log{\Big{(}{{2\pi l_{\rm P}}\over{\theta \lambda}}\Big{)}}\Big{]}\Big{\}}, \eqno(5)$$where the right-hand-side involves integration by parts, split at $R(\lambda/D)^\rho$. For a telescope of diameter $D$, the {$\Phi_{\rm R}$} portion includes all phase dispersions up to the instrumental resolution limit, or $R=1.22$ and $\rho=1$ for a perfect, circular and unobstructed aperture, set by diffraction. Above that the foam-blurred PSF still grows towards high energy, but more slowly where hemmed-in by $\theta$.

This asymptotic PSF behaviour should feel familar to ground-based optical/near-infrared astronomers: blueward of the horizon-crossing wavelength the quantum-foam-induced PSF can still be larger than the telescope resolution and less than the FoV, but now has four self-similar levels of blurring. The narrowest is a sharp PSF core for those photons that all fell inside the instrument resolution, and three progressively broader layers for those falling either outside the mean, characteristic Theta angle or its FoV, and eventually the horizon itself. And so Theta is analogous to ``seeing'' caused by atmospheric distortions when viewed from a ground-based site, although that is only about 1 arcsec across or less from the best sites; but akin to quantum foam, occurs due to fluctuations in index-of-refraction in turbulent air along the telescope beam, falling well below the resolution limit of any real telescope blueward of ultraviolet light.

\subsection{The Measured Point-Spread Function for {\it Fermi}}

\begin{figure}
\hspace{4.5cm}\includegraphics[scale=1.00]{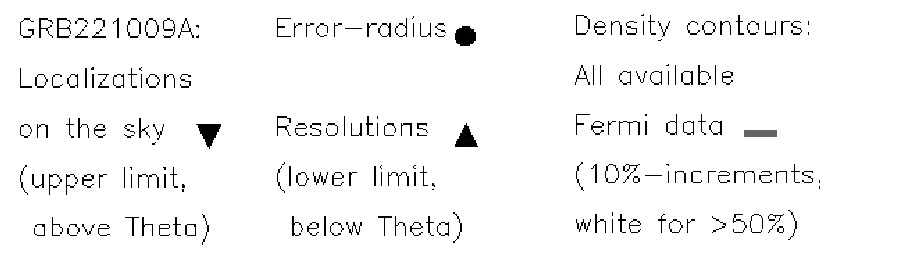}\\
\includegraphics[scale=0.40]{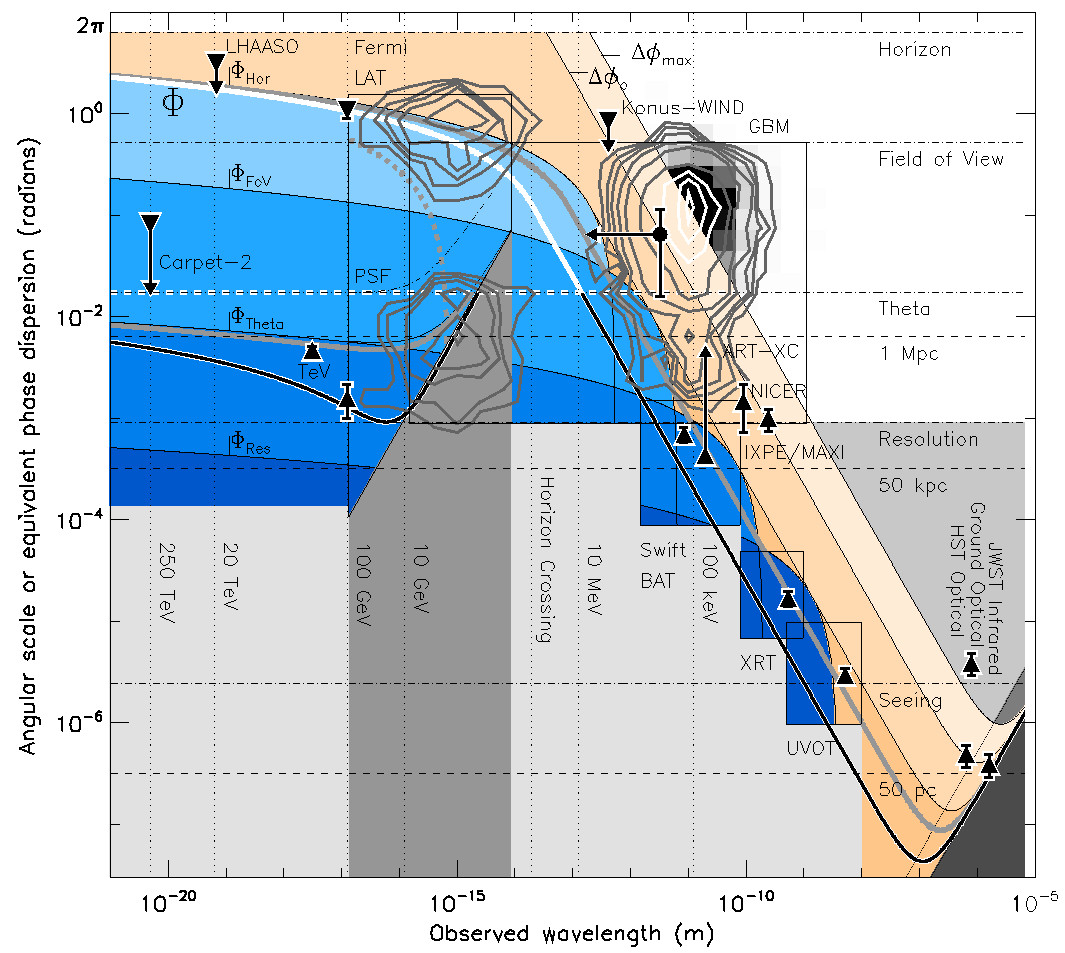}
\caption{Measurements of GRB221009A, reproduced from \cite{Steinbring2023}. Density contours are of all available, archival GRB data from {\it Fermi} LAT (roll-angles and resolutions) or GBM (error-radii) shown in 10\% increments, white above 50\% for GBM. Note how the broadest case of a real instrumental PSF (white curve; equation 5) nicely matches the roll angle/zenith angle for the {\it Fermi}-LAT instrument (shown for $\alpha=0.650$, $z=0.151$, and black-on-white where below Theta; grey-curve: same, $z=1.41$) or the blueward edge of the error radius for the GBM ($\Phi_{\rm FoV}$). The lower limit where blurring must necessarily rise above the nominal resolution ($\Phi_{\rm Res}$) then occurs at $\alpha=0.735$, also shown as a black-on-white curve, where it is less than Theta. In between, scaling from an upper limit of $\Phi_{\rm Hor}$ to $\Phi_{\rm Theta}$, the last term in equation 5 would follow the ratio $1+2\pi/(1.22\times\theta)$, where $\theta$ is $1^\circ$. And so, the half-way point for $\alpha$ (i.e., just scaled instead by the ratio of the resolution to the horizon) happens to coincide with $\alpha=0.667$, which is the holographic value favored by QG models; the associated horizon-crossing angle is also indicated. This limiting behaviour - that an instrument records only the ensemble of wavelength-dependent scattering cases more than its resolution-limit and less than its FoV - is something like the effect of seeing from the ground, and leads to the smooth scaling for the average size of the bluest $\gamma$-ray sources (plotted as a dashed gray curve for the average redshift $z=1.41$). This regime will be scrutinized more carefully in the discussion to follow.}
\label{figure_wide}
\end{figure}

Recall that GBM and LAT cannot resolve down to their diffraction limits; the LAT is actually a pair-production instrument sensitive to $\gamma$-rays through resultant electron/positron tracks, using scintillometers to reject background events, and a separate calorimeter recovers the incident $\gamma$-ray energy. That is why it has such wide-field view: sensing photons from almost any direction. But like an imager the detected area on the sky of sensed $\gamma$-rays for a particular source does effectively sample over what constitutes its resolution limit and an averaged fit of those for their mean energy can be considered the effective instrumental PSF of the telescope. That was measured in-orbit for LAT using images of a large sample of AGNs, and gives the scaling relation $${\rm PSF}\propto\sqrt{[(C_0 E/100)^{-\rho}]^2 + C_1^2}, \eqno(6)$$where $E$ is energy in MeV; $C_0=3.5^\circ$ and $C_1=0.15^\circ$, and $\rho=0.8$ is the power-law slope \citep{Ackermann2013}. In Figure~\ref{figure_wide} this is plotted as a dashed curve, scaled to Theta, which happens to be near the mid-point of {\it Fermi} LAT/GBM resolutions; the mean PSF of those instruments.

\section{Observations}

The available GRB data are now compared to the theoretical PSF derived in the previous section. And there is one GRB that is particularly special, as it was the brightest, highest-energy such event ever observed: GRB221009A. All of those data are already presented elsewhere, and summarized in \cite{Steinbring2023}, but reproduced here. The intent will be to emphasize, particularly with {\it Fermi} GBM and LAT data, the remarkable adherence to the model PSF and characteristic Theta described above, along with the in-orbit measured PSF; the case is reiterated that this may simply be due to previous studies incorrectly accounting for instrumental effects, and so failing to detect the blur.

\subsection{The Special Case of GRB221009A}

On 9 October 2022 at 13:16:59 UT GRB221009A triggered the {\it Fermi} GBM within its FoV of $35^\circ$ \citep{Veres2023}, locating it inside an error-radius of $3.71^\circ$ (90\% confidence) at peak energy 375 keV, simultaneously with LAT \citep{Bissaldi2023}. Those instrumental entendues are outlined in Figure~\ref{figure_wide}: the latter has a resolution of $5^\circ$ at 30 MeV or 1.5 arcmin at 60 GeV, and together with GBM has found over 3390 GRBs (median-$z$: 1.41; highest-$z$: 4.61) from 100 MeV to 100 GeV since launch in 2008. For comparison, contour plots of all localization data available from the High-Energy Astrophysics Science Research Archive: https://heasarc.gsfc.nasa.gov/ (as accessed on 1 November 2022) are plotted using the peak detected wavelength for each, scaled by $1/(2\pi{\rm c}\hbar)$.

Once triggered, finding GRB221009A with LAT at such high energy restricts Planck-scale induced blurring to be {\it weak}, as photons were detected at 100 GeV within 1 radian at 397.7 GeV; so not all could have been scattered to the horizon. And the first-ever detection of at least one higher-energy photon can also be inferred by Carpet-2 via cosmic-ray air-shower angle-of-arrival (AoA) at 251 TeV. That instrument has an all-sky FoV (essentially $2\pi$) and resolution of $4.7^\circ$ (90\% confidence), with a minimum instead set by the angular distance to the optical transient position (down-pointing arrow in Figure~\ref{figure_wide}). But that must set a critical angle near $1^\circ$, because if quantum-foam-induced blur is present at this wavelength, it requires that here (and at all longer wavelengths) some wavefronts phase-dispersed less.

How {\it strong} foam-induced blurring can be was also restricted by many observations: shown in Figure~\ref{figure_wide} are {\it Swift} satellite detection with the Burst-Alert Telescope (BAT) X-Ray Telescope (XRT), and Ultra-violet Optical Telescope (UVOT); black symbols indicate the positional accuracy, both roll-angle/zenith-angle or error-radius (above Theta) and localization/resolution (below Theta). The most-stringent of these is at 18 TeV by the Large High Altitude Air Shower Observatory (LHAASO), which allows only a broad sky localization much like LAT roll-angle; but there are many more: the space-based instruments Konus-{\it WIND} (KW) and {\it Mikhail Pavlinsky} Astronomical Roentgen Telescope X-ray Concentrator (ART-XC), the Neutron star Interior Composition Explorer (NICER) and Monitor of All-sky X-ray Imager (MAXI), the Imaging X-ray Polarimetry Explorer (IXPE) along with imaging from HST, the {\it James Webb Space Telescope} (JWST) and a redshift obtained from ground-based optical telescope spectroscopy. 

Thus, having located that object on the sky to within a degree in higher-energy $\gamma$-rays than seen before, and then later being able to identify its host galaxy places lower and upper limits on blurring attributable to spacetime. And nicely, equation 5 with $\alpha=0.677$ indeed indicates that a gamma-ray telescope sensitive beyond 10 GeV should have seen those photons from almost anywhere on the sky within its FoV of about 90 degrees, but most photons would fall within 1 degree, and an image could have a sharpest resolution of roughly 1 arcminute. In fact, in the next section it will be shown that the observed {\it Fermi} PSF scaling of equation 6 can be explained by this, which simultaneously resolves the tension between GRBs being detected at extremely high energies, and then later being precisely identifiable within galaxies at optical/near-infrared wavelengths, while being consistent with previous observations of quasars at all energies. 

\subsection{Seeing a Halo Attributable to HQF-Blurring}

\begin{figure}
\hspace{4.5cm}\includegraphics[scale=0.40]{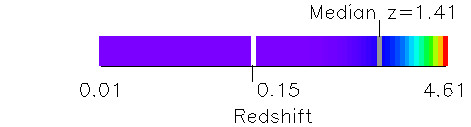}\\
\includegraphics[scale=0.40]{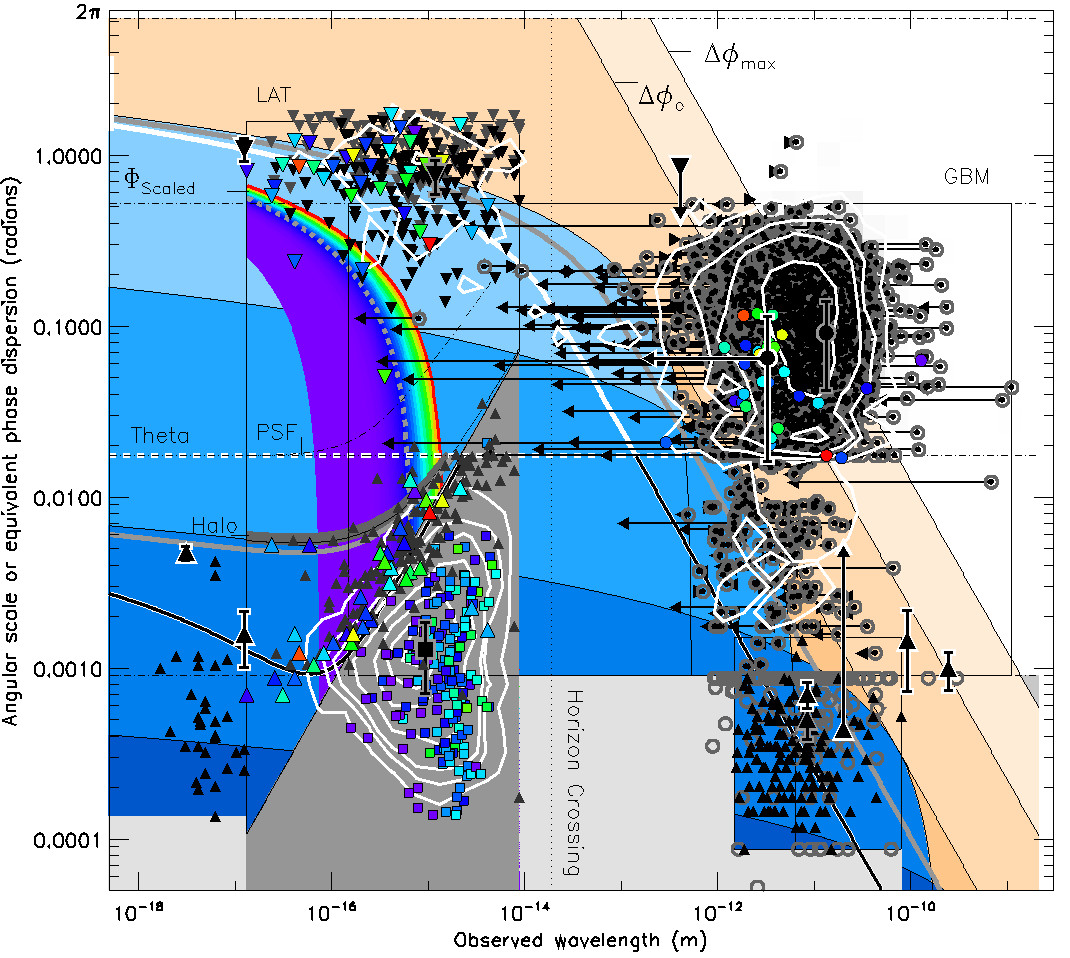}
\caption{Same as Figure~\ref{figure_wide}, now with all individual GRB measurements shown; reproduced from \cite{Steinbring2023} and restricted to $\gamma$-ray wavelengths to show the highest-energy LAT photons detected. The down-pointing triangles are all archival roll-angles (black) and zenith angles (grey) available from the {\it Fermi} Archive as of 1 November 2022; up-pointing triangles are their corresponding resolutions. Some left-pointing triangles indicate the highest-energy detections from GBM; grey circles indicate the mean energy. At best PSF-sharpness, the open squares are the cataloged {\it Fermi} LAT resolutions measured for AGNs (each within a narrow energy range) and colour-coded by its redshift, where known; the PSF scaling behaviour described in Figure~\ref{figure_wide} is shown colour-coded in the same way. White density contours outline the full sample in 10\% increments. To the left, at higher energies than ever detected for AGNs, are TeV sources (black, up-pointing triangles) from the TeVCat catalog (complete as accessed on 1 November 2022 from http://tevcat.uchicago.edu). Model curves are as described before, but here light-grey: $\alpha=0.650$, $z=1.41$; black-on-white: $\alpha=0.735$, $z=0.151$. The most remarkable aspect of this plot is the agreement of the best-resolved highest-energy LAT GRBs with the dark-grey curve labelled ``Halo,'' as this would be the expected result, limited by quantum foam for $\alpha=0.667$.}
\label{figure_both}
\end{figure}

Scrutiny can now turn to how well the theoretical foam-induced $\Phi$ compares to the {\it Fermi} data in detail. Notice how the expected upper-limit of blurring matches the mean distribution of roll-angles (including GRB221009A) in Figure~\ref{figure_both}, seen at upper-left. And below that, see how the model of quantum-foam-induced blur plus LAT instrument resolution limit (thick grey curve labelled ``Halo'') nicely matches the poorest-resolved GRBs. And agreement continutes to where they turn off in approaching angle Theta at the lowest, poorest resolved LAT energies. This can explain how those ``tail off'' for the largest image sizes at the longest wavelengths detectable with LAT, despite what one would otherwise expect to worsen with instrumental resolution (the slope of the dark-grey shaded region). The GRBs are still point-sources here, but the upper extent of this measured PSF-scaling is actually from a catalogue of observed AGNs (Abdollahi et al. 2020; 4050 samples, mean-$z$: 0.967); their distribution is shown as white density contours (1\%, 5\%, 10\%, 25\%, 50\%, 75\%, 90\%; and, where available, filled squares color-coded by redshift using scale shown above. Equation 5 anticipates this, because when these are instead true point sources of a given redshift blurred only by foam, the last term in that equation allows that each can accumulate any image size smaller than Theta, down to the resolution limit, despite that being less than the mean resolution at that energy.

Clearly, LAT images are not limited by diffraction but scale with it as predicted by foam-induced blurring, which sets a resolution floor indicated here by the lower dot-dashed horizontal line. Note that for the highest energies detectable with LAT (shortest wavelengths) the weakest possible $\Phi_{\rm Theta}$ (black on white curve, $\alpha=0.735$, $z=0.151$) is still consistent with localization of GRB221009A. And anywhere in between, the average redshift (dashed curve) agrees with the LAT and GBM measured energies as well: either the down- and left-pointing arrows for any of those sources. It is expected that they could lie anywhere within the sensitivty of LAT and GBM, but blurred by holographic-QG foam ($\alpha=0.667$) none should stray blueward of this smooth, redshift-scaled demarcation (colour-coded in the same way as the symbols). Assuming that one-sigma from the mean angular localization of the sample is a reasonable estimate of uncertainty, that is true, and accommodates two possible outliers (one dark blue and one green down-pointing triangle) in the LAT GRB roll-angle sample.  It is indeed an excellent fit to all of the data.

\vspace{3 cm}

\section{Summary and Conclusions}\label{conclusions}

To summarize the first part of our talk on the theoretical underpinnings of HQF, we have employed various methods to show that the uncertainty $\delta l$ in the measurement of a distance $l$ scales as $l^{1/3}$ as demanded by the holographic principle. And we show that $\delta l$ would scale as $l^{1/2}$ if there were only ordinary matter obeying Bose or Fermi statistics in the Universe.  In other words, the existence of a dark sector is a prediction of HQF! Next we show that quanta of the dark sector have to obey an "exotic" statistics, viz. the infinite statistics. (Could this be a reason why we have not been able to detect conclusively dark matter "particles"?) Furthermore it is known that any theory of infinite statistics is fundamentally non-local (albeit consistent with Lorentz and CPT invariance). As far as we know, such a complete field-theoretical description of infinite statistics is not available at present.  (And thus we have to start from scratch, perhaps with an effective field theory?)

The rest of the first part deals with cosmology --- more specifically, the holographic foam cosmology: properties of dark energy, dark matter, and cosmic inflation in the early Universe.  We conclude with a heuristic argument that, of all the quantum foam models (parametrized by $\delta l \sim l^{1 - \alpha} l_P^{\alpha}$), Kolmogorov-type scaling in turbulence is consistent only with HQF ($\alpha = 2/3$), verifying Wheeler's picture of spacetime foam: spacetime, probed at small scales, is a turbulent froth. 

The second part of this talk deals with the observational evidence of HQF. Before we give a summary for that part, we feel we would be remiss if we fail to mention another possible way to detect quantum foam, viz., with laser-based interferometers \citep{Amelino-Camelia1999, Ng2000}. For an interferometer with bandwidth centered at frequency $f$, the relevant length scale characteristic of the noise due to space-time foam is given by $l_P^{2/3} (c/f)^{1/3}$. This uncertainty manifests itself as a displacement noise (in addition to noises from other sources) that infests the interferometers. The hope is that modern gravitational-wave interferometers, through future refinements, may reach displacement noise level low enough to test a subset of the space-time foam models.


Finally let us summarize the second part of our talk on the observational evidence of HQF. 
What a telescope will see when viewing a quantum-foam blurred source depends on its field of regard, its resolution and the energy range. A model PSF, appropriate for pointlike GRBs is derivable. And accounting for realistic entendues of telescopes is able to explain both the broad multi-degree angular spread in localization of GRBs on the sky in high-energy $\gamma$-rays and X-rays, and how these can later still be identified at sub-arcsecond scales within their source galaxy via optical/near-infrared spectroscopy. Averaging over the sensitivity ranges of real $\gamma$-ray telescopes, as any imaging system does, accounts for the consequences of some photons falling outside the telescope FoV (consistent with these being seen at all by LHASSO, Carpet-2 at TeV energies) and some within the resolution and energy-sensitivity of an instrument (explaining why optical/near-infrared telescopes such as HST and JWST see no strong evidence of foam). 

Quantum-foam blurring is a particularly good fit for data obtained with {\it Fermi} GBM and LAT, especially if $\alpha=0.667$, and suggests HQF could be responsible for the resolution-limiting behavour of its PSF, which is of direct interest to $\gamma$-ray astronomy more generally. The recent GRB 221009A was the brightest one ever observed, and visible across a huge wavelength range. Importantly, its highest-energy photons should be the most-scattered by spacetime-foam, with dispersions of many degrees.  Holographic foam-induced blurring sets a mean angular size $\Phi_{\rm Theta}$ of any GRB to be close to $1^\circ$ near 250 TeV, the highest energy localization of GRB221009A. Interpreted that way, GRB221009A provided the first clear observational evidence that spacetime is not smooth, and is consistent with HQF, as is all previous imaging of galaxies, AGNs, and other available GRB localizations known.


\begin{thebibliography}{99}

\bibitem[Abdollahi et al.(2020)]{Abdollahi2020} Abdollahi, S., Acero, F., Ackermann, M., Ajello, M.  2020, Astrophs. Journal Supp. Series, 247 33
\bibitem[Ackermann et al.(2013)]{Ackermann2013} Ackermann, M., Ajello, M., Allafort, A., et al. 2013c, Astrophys. Journal, 765, 54
\bibitem[Arzano et al. (2007)]{Arzano2007}  Arzano, M., Kephart, T.W., \& Ng, Y.J.  2007, Phys. Lett. B, 649, 243
\bibitem[Bissaldi et al.(2023)]{Bissaldi2023} Bissaldi, E., Omodei, N., Kerr, M. et al. 2023, GRB Coordinates Network, Circular Service, 32637
\bibitem[Amelino-Camelia (1999)]{Amelino-Camelia1999}  Amelino-Camelia, G.  1999, Nature, 398, 216
\bibitem[Amelino-Camelia(2013)]{Amelino-Camelia2013} Amelino-Camelia, G. 2013, Living Reviews in Relativity, Volume 16, article number 5
\bibitem[Carlip(2023)]{Carlip2023} Carlip, S. 2023, Rep. Prog. Phys. 86 066001
\bibitem[Christiansen et al.(2011)]{Christiansen2011} Christiansen, W.A., Ng, Y.J., Floyd, D.J.E., \& Perlman, E.S. 2011, Physical Review D, 83, 84003
\bibitem[Greenberg (1990)]{Greenberg1990}  Greenberg, O.W. 1990, Phys. Rev. Lett., 64, 705
\bibitem[Ho et al. (2012)]{Ho2012}  Ho, C.M., Minic, D., \& Ng, Y. J. 2012, Phys. Rev. D , 85, 104033
\bibitem[Jejjala et al. (2008)] {Jejjala2008}  Jejjala, V., Minic, D., Ng, Y. J., \& Tze, C. H. 2008, Class. Quant. Grav., 25, 225012
\bibitem[Karolyhazy (1966)] {Karolyhazy1966}  Karolyhazy, F. 1966,  Il Nuovo Cimento, A 42, 390
\bibitem[Kolmogorov (1941)]{Kolmogorov1941}  Kolmogorov, A. N. 1941, Dokl. Akad. Nauk SSSR, 30, 299
\bibitem[Lieu \& Hillman(2003)]{Lieu2003} Lieu, R. \& Hillman, L.W. 2003, Astrophys. Journal Lett., 585, L77 
\bibitem[Lloyd \& Ng (2004)]{Lloyd2004}  Lloyd, S. \& Ng, Y.J. 2004, Sci. Am., 291 (5), 52
\bibitem[Margolus \& Levitin (1998)]{Margolus1998}  Margolus, N. \& Levitin, L.B. 1998, Physica D, 120, 188
\bibitem[Milgrom (1983)]{Milgrom1983} Milgrom, M. 1983, Astrophys. J., 270, 365
\bibitem[Ng \& van Dam (1994)]{Ng1994} Ng, Y.J. \& van Dam, H. 1994, Mod. Phys. Lett. A , 9, 335
\bibitem[Ng \& van Dam (2000)]{Ng2000} Ng, Y.J. \& van Dam, H. 2000, Found. Phys. , 30 (5), 795
\bibitem[Ng, Christiansen, \& van Dam(2003)]{Ng2003} Ng, Y.J., Christiansen, W.A., \& van Dam, H. 2003, Astrophs. Journal Lett., 591, L87
\bibitem[Ng (2007)]{Ng2007}  Ng, Y.J. 2007, Phys. Lett. B, 657, 10
\bibitem[Ng (2021)]{Ng2021}  Ng, Y J  2021, Symmetry, 13, 435
\bibitem[Ng \& Perlman(2022)]{Ng2022} Ng, Y.J. \& Perlman, E. 2022, Universe, 8, 382
\bibitem[Ng \& Steinbring(2025)]{Ng2025} Ng, Y.J. \& Steinbring, E. 2025, International Journal of Modern Physics D, 2544018
\bibitem[Perlman et al.(2011)]{Perlman2011} Perlman, E.S., Ng, Y.J., Floyd, D.J.E., \& Christiansen, W.A. 2011, Aston. and Astrophys., 535, L9
\bibitem[Perlman et al.(2015)]{Perlman2015} Perlman, E.S., Rappaport, S.A., Christiansen, W.A., Ng, Y.J., DeVore, J., \& Pooley, D. 2015, Astrophys. Journal, 805, 10
\bibitem[Ragazzoni, Turatto, \& Gaessler(2003)]{Ragazzoni2003} Ragazzoni, R., Turatto, M., \& Gaessler, W. 2003, Astrophys. Journal Lett., 587, L1
\bibitem[Sorkin (1997)] {Sorkin1997}  Sorkin, R. D. 1997, Int. J. Th. Phys., 36, 2759
\bibitem[Steinbring(2007)]{Steinbring2007} Steinbring, E. 2007, Astrophys. Journal, 655, 714
\bibitem[Steinbring(2015)]{Steinbring2015} Steinbring, E. 2015, Astrophys. Journal, 802, 38
\bibitem[Steinbring(2016)]{Steinbring2016} Steinbring, E. 2016, IAU Symposium Conf. Series, 319, 54
\bibitem[Steinbring(2023)]{Steinbring2023} Steinbring, E. 2023, Galaxies, 11(6), 115
\bibitem[Steinbring(2023)]{Steinbring2023b} Steinbring, E. 2023, Astrophysics Source Code Library, record ascl:2311.003
\bibitem[Steinbring(2025)]{Steinbring2025} Steinbring, E. 2025, Journal of Physics Conference Series, 2987 012016
\bibitem[Susskind (1995)]{Susskind1995}  Susskind, L. 1995, J. Math. Phys. (N.Y.) , 36, 6377
\bibitem[Tamburini et al.(2011)]{Tamburini2011} Tamburini, F., Cuofano, C., Della Valle, M., \& Gilmozzi, R. 2011, Astron. and Astrophys., 533, A71
\bibitem['t Hooft (1993)]{'t Hooft1993}  't Hooft, G. 1993, Salamfestschrift, World Scientific, 284
\bibitem[Veres et al.(2023)]{Veres2023} Veres, P., Burns, E., Bissaldi, E., Lesage, S. et al. 2023, GRB Coordinates Network, Circular Service, 32636
\bibitem[Wheeler(1957)]{Wheeler1957} Wheeler, J.A. 1957, Annals of Physics, 2, 604

\end{thebibliography}
\end{document}